\documentclass[showpacs,prl,aps,floatfix,reprint]{revtex4-1}

\usepackage{amsmath}
\usepackage{amsfonts}
\usepackage{graphicx}
\usepackage{bm}

\newcommand{\ix}[1]{\text{ #1}}
\newcommand{\mc}[1]{\ensuremath{\mathcal{#1}}}
\newcommand{\Spm}[1]{\mathcal{S}_{#1 }^{\,\pm}}
\newcommand{\Sm}[1]{\mathcal{S}_{#1 }^{\,-}}
\newcommand{\Sp}[1]{\mathcal{S}_{#1 }^{\,+}}
\newcommand{\szn}[1]{\mc S_{#1}^z}
\newcommand{\mean}[1]{\ensuremath{ \langle\,#1\, \rangle}}

\allowdisplaybreaks[3]

\begin{document}

\title{Formation of supermodes in atom-microcavity chains}

\author{Sandra Isabelle \surname{Schmid}}

\author{J\"org \surname{Evers}}

\affiliation{Max-Planck-Institut f\"ur Kernphysik, Saupfercheckweg 1, D-69117
Heidelberg, Germany} 


\date{\today}

\begin{abstract}
A chain of atom-microcavity systems coupled by a common fiber is considered.  We analyze the formation of supermodes and focus on  the dependence of this effect on the chain geometry and the number of atom-cavity subsystems. We show that the significance of supermodes to the transmission increases with the number of atom-cavity subsystems.  We identify spectral ranges in which the chain geometry decides whether supermodes are formed, and ranges which are insensitive to the geometry.  Furthermore, we show that the reflection signal allows to identify which cavities couple to atoms, which is a crucial information in experimental realizations of longer atom-cavity chain systems.
\end{abstract}

\maketitle

Whispering gallery microresonators have gained tremendous interest over the past few years due to their outstanding properties~\cite{N5}. Particularly promising photonic systems arise if quantum systems such as atoms are coupled to the evanescent field of microresonators.
\begin{figure*}
\includegraphics[width=15cm]{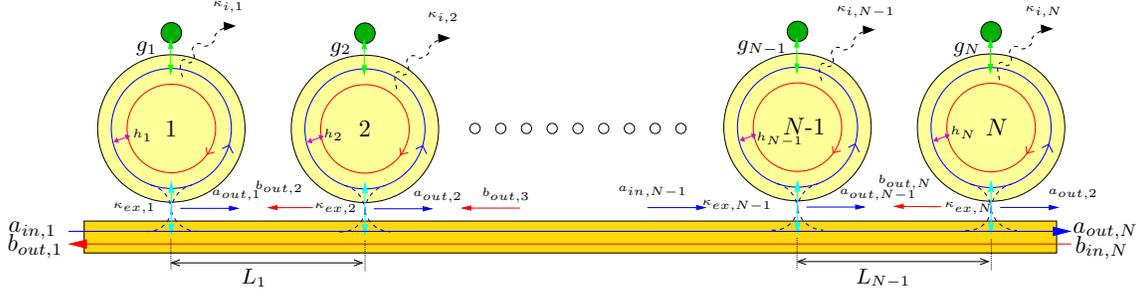}
\caption{Chain of $N$ cavity-atom subsystems connected bi-directionally by a waveguide. }
\label{skiz}
\end{figure*}
Already the simplest case of a single particle coupled to a single microcavity has led to a number of fascinating proposals and experiments. For example, strong coupling of the microcavity was demonstrated to atoms~\cite{strong} and quantum dots~\cite{painter}, a single photon turnstile was realized in~\cite{turnstile}, and the complex interplay of counterpropagating resonator modes already due to a classical particle was observed in~\cite{PhysRevLett.99.173603}.
Small chains of atom-cavity systems have been analyzed as well. For example, in~\cite{zoller}, two coupled atom-cavity systems are considered for applications in quantum networks~\cite{quantint}. But in this work, light can propagate only in one direction through the system, and backward couplings and thus supermodes between the subsystems are neglected.
This restriction was lifted in~\cite{zang2010}, in which two atom-cavity systems coupled to a fiber are considered including the scattering between the cavities. However, only a single cavity mode instead of a pair of modes was considered in each resonator. The non-coupling eigenmode of the resonator found, e.g., in~\cite{strong}, which can have strong influence on the system's optical properties, is not considered. Furthermore,  only the transmission was analyzed, and cases with more than two atom-cavity systems, the reflection,  and the formation of supermodes were not studied.
Meanwhile, recent experiments indicate that a realization of chains of coupled atom-cavity systems is within reach. For example, next to the single atom-cavity systems~\cite{turnstile,strong}, it was also shown that atoms can be trapped by the evanescent field of a tapered fiber~\cite{Rauschenbeutel2010,PhysRevA.70.063403}. Furthermore, real-time detection and feedback to monitor single atoms near a microresonator was recently achieved, which is an important step towards larger networks of atom-cavity systems~\cite{alton2010}.

Motivated by this, here, we investigate  a chain of $N$ coupled cavity-atom subsystems connected via a waveguide and probed by a weak input field, see Fig.~\ref{skiz}. We mainly study the formation of supermodes, which can arise due to the scattering of light between the different subsystems. To identify these modes, we define a ``superness'' measure, which is given by the difference in transmission for the complete system relative to the transmission to a corresponding system of independent cavities without backward coupling. We find that the formation of supermodes crucially depends on the length of the chain, and the relative distances between the subsystems. While the overall transmission decreases with increasing number of atom-cavity subsystems in the chain, the relative contribution of the supermodes increases. As in experiments it can be difficult to achieve simultaneous coupling of all resonators to individual atoms, we also show that the atom-cavity coupling configuration of the chain can be determined from the reflected light, whereas the transmitted light does not contain this information. In this sense, the reflection and transmission contain complementary information.

Our system consists of a chain of $N$ microresonator-atom systems coupled via a fiber,  see Fig.~\ref{skiz}. Due to the intercavity distance direct couplings between cavities are neglected. Each resonator $n\in\{1,\dots,N\}$ is modeled by a pair of counterpropagating modes described by the annihilation operators $a_n, b_n$, with scattering between the modes of rate $h_n$. The coupling strength to the fiber is  $\kappa_{ex,n}$, and each cavity has an internal loss rate $\kappa_{i,n}$ such that the total loss rate is $\kappa_i = \kappa_{i,n}+\kappa_{ex,n}$. The detuning between the resonator frequency $\omega_{cav,n}$ and the input field frequency $\omega_L$ is $\delta_n = \omega_{cav,n} - \omega_L$. The distance between the coupling points of two resonators $n$ and $n+1$ to the fiber is $L_n$.  
The atom at resonator $n$ is modeled as two-level systems with resonance frequency $\omega_{at,n}$, decay rate $\gamma_n$, atomic raising (lowering) operators $\Spm{n}$ and detuning $\Delta_n = \omega_{at,n}-\omega_L$ to the incident light. The position-dependent coupling constants of the atom to modes $a_n$ and $b_n$ are $g_{n,a}$ and $g_{n,b}$, respectively.
Due to the scattering $h_n$ it is convenient to introduce normal modes 
$\mc A_{n}=(a_{n}+b_{n})/\sqrt 2$ and $\mc B_{n}=(a_{n}-b_{n})/\sqrt 2)$. Depending on the position of the atom along the circumference, it can selectively couple to either of the modes, or to both at the same time~\cite{turnstile,strong}.
With these definitions the Hamiltonian reads
\begin{align}
\mc H_N=&\sum_{n=1}^N-\hbar\Delta_n \Sm{n} \Sp{n} \nonumber\\
&+(\delta_n+|h_n|) \mc A_n^+ \mc A_n+\hbar (\delta_n-|h_n|) \mc B_n^+ \mc B_n\nonumber\\
&+i\hbar\sqrt{2\kappa_{ex,n}}\left(\mc A_{in,n}\mc A_n^\dagger+ \mc B_{in,n}\mc B_n^\dagger+ \ix{H. c.}\right)\nonumber\\
&+ \hbar \left(g_{\mc A_n}\mc A_n^\dagger\Sm{n} -ig_{\mc B_n}\mc B_n^\dagger \Sm{n} +\ix{H. c.} \right)\:.
\end{align}
where we define $\szn{n}=[\Sp{n},\Sm{n}]$, and input fields to resonator $n$  are denoted $\{\mc A_{in,n},\mc B_{in,n}\}$.
For our calculations we use a semiclassical treatment and replace operators with their respective expectation values~\cite{painter}.  
The coupling of the different cavities arises since the input flux of cavity $n$ depends on the outputs of the neighboring cavities via
$a_{in,n}=a_{out,n-1}\exp(ikL_{n-1})$ and $b_{in,n}=b_{out,n+1}\exp(ikL_{n+1})$. Then together with the input-output relations~\cite{Gardiner1985} $a_{out,n}=-a_{in,n}+\sqrt{2\kappa_{ex,n}} a_n$, 
all input and output fluxes can be determined.


In general, solving the coupled system for $N$ cavities is a demanding task due to the large dimension of the Hilbert space. One approach to solve a system of two coupled atom-cavity systems  was presented in~\cite{zang2010}, based on a real space wavefunction approach for a single photon wave packet~\cite{PhysRevA.79.023837}. However, it is well known that resonators without atoms can be described by transfer matrices relating inputs and outputs of a single resonator~\cite{Xiao2008}.  We found that a related approach using transfer matrices $\mathcal{M}_n$ is also possible for resonators coupled to an atom in certain parameter regimes. Thus, each atom-cavity system can be solved separately, avoiding the large Hilbert space of the combined $N$-cavity system. We verified using exact calculations of smaller multiple cavity systems that this approach is possible as long as the atoms are far from saturation. Then, the $n$th cavity-atom system can be described via the frequency-dependent transmission $t_n$ and reflection $r_n$ determined by
$a_{out,sc}=t_n(\Delta_n)\cdot a_{in,sc}+r_n(\Delta_n)\cdot b_{in,sc}$ and $b_{out,sc}=r_n(\Delta_n)\cdot a_{in,sc}+t_n(\Delta_n)\cdot b_{in,sc}$, out of which the transfer matrix  $\mathcal{M}_n$ can be formed. In this formalism, the optical path between the cavities is characterized by a diagonal matrix $\mc M_{\phi_n}$ with eigenvalues $\exp[i\phi_n]$ and $\exp[-i\phi_n]$, where the phase angles $\phi_n$ are defined by the intercavity distances $L_n$ as $\phi_n=2\pi L_n/\lambda$ where $\lambda$ is the wavelength of the incident light.
The total system is then governed by the matrix $\mc M_{total}=\mc M_N\cdot\mc M_{\phi_{N-1}}\cdot\mc M_{N-1}\cdots\mc M_{2}\cdot\mc M_{\phi_1}\cdot\mc M_1$,  which connects the input fluxes $a_{in,1}$ and $b_{in,N}$ with the output fluxes 
$a_{out,N}$ and $b_{out,1}$.


We now turn to a discussion of our observables. Assuming driving of the system from the left side only ($b_{N,in}=0$),  the transmission $T$ and reflection $R$ are given by 
$T=|\mean{a_{out,N}}|^2/|a_{in,1}|^2$ and 
$R=|\mean{b_{out,1}}|^2/|a_{in,1}|^2$.
%
Next, we are interested in studying the formation of supermodes. These are modes which receive nontrivial contribution from multiple cavities, thus having properties which go beyond the combination of the properties of the individual subsystems. Hence, we define the {\it ``superness''} of a mode, $\Delta T=T-T_{ind,N}$, which is  the difference between the transmission obtained for the full $N$-cavity system with backward coupling to the transmission for a chain of $N$ independent cavities $T_{ind,N}=T_1\cdot T_2\cdots T_N$ without backward coupling.

For our numerical results we assume that the atoms are located such that they couple to the modes $\mc B_n$ and set $g_{\mc A_n}=0$. Note that the normal mode $\mc A$ still contributes to the optical properties as it couples to the incident probe beam. Furthermore, identical spontaneous emission rates for the atoms $\gamma_n=\gamma$ are assumed.

\begin{figure}[t]
\includegraphics[width=8cm]{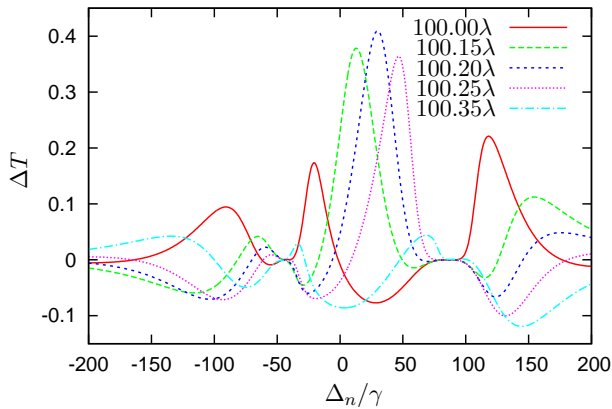}
\caption{\label{Th5}(Color Online) Formation of supermodes indicated by large modification $\Delta_T$ of transmission compared to a chain of independent atom-cavity systems. Parameters are $N=2$, $h_n=50\gamma$, $g_{\mc B_n}=70\gamma$, $L_{tot}=200.3\lambda$ and intercavity distances $L_1=100.0\lambda$, $L_1=100.15\lambda$, $L_1=100.25\lambda$, and $L_1=100.35\lambda$.}
\end{figure}

We start with $N=2$. Figure~\ref{Th5} depicts our results for $\Delta T$ for two subsystems, $h_n=50\gamma$ and different distances $L_n$. We observe that for specific detunings the transmission behavior crucially depends on the intercavity distance whereas for other detunings the systems behaves almost as a chain of independent cavity-atom systems. In particular, at $\Delta_n\approx 95\gamma$ and $\Delta_n\approx -50\gamma$, the transmission of the full system is equal to the combination of the individual systems for all distances. This can be explained by noting that for these  detunings, either $|t_n|$ (for $\Delta_n\approx 95\gamma$) or  $|r_n|$ (for $\Delta_n\approx -50\gamma$) is small. This leads to a vanishing backcoupling between the cavities such that the system is similar to the uncoupled case. For $\Delta_n\approx -50\gamma$ the population of the upper level  of both the atoms is comparably high and the presence of the atom supresses the reflection which results in $|r_n|\approx 0$. 
For $\Delta_n\approx 95\gamma$ almost all incoming light is reflected at the first subsystem and thus the cavity modes and the excited state of the atom next to cavity two are hardly populated. The low transmittivity and reflectivity are achieved since the respective detunings correspond to the eigenvalues of the Hamiltonian for a single subsystem. 
In contrast, for other detunings, pronounced resonances in the ``superness'' $\Delta T$ can be observed at certain distances. These resonances exceed $\Delta T\approx 0.4$ for optimum parameters, demonstrating clearly the relevance of the backscattering between the cavities. For these detunings $|t_n|\approx |r_n|$ and thus the energy exchange in both directions is enhanced. Interestingly, for the detunings with high $\Delta T$, the modes $\mc B_n$ coupled to the atoms are hardly populated in both the cavities, whereas the non-coupling modes $\mc A_n$ are highly populated, demonstrating the significance of the non-coupling normal mode.

\begin{figure}[t]
\includegraphics[width=8cm]{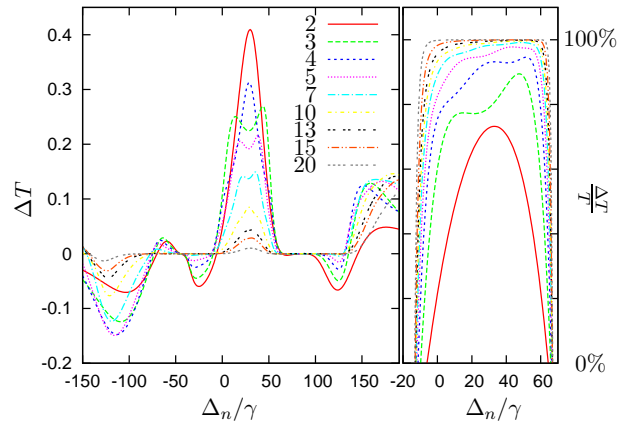}
\caption{\label{Nxsuper}(Color online) Dependence of the supermode contribution $\Delta T$ on the chain length $N$. The left panel shows the absolute ``superness`` $\Delta T$, the right panel the relative value $\Delta T / T$. The parameters are as in Fig.~\ref{Th5}, with $L_n=100.2\lambda$.}
\end{figure}

We now turn to larger arrays of atom-cavity systems. The dependence of the supermode identified in Fig.~\ref{Th5} on the chain length from $N=2$ up to $N=20$ subsystems is shown in Fig.~\ref{Nxsuper}. We found that in absolute terms, the superness $\Delta T$ reduces with increasing $N$ until it almost vanishes for $N=20$. This reduction can be traced back to the overall reduction in transmission $T$ for increasing $N$, since each resonator leads to a certain amount of loss. In contrast, the ''relative superness`` $\Delta T / T$ increases with $N$, as shown in the right panel of Fig.~\ref{Nxsuper}. The reason is that the reduction in transmission becomes stronger with increasing $N$ for independent resonators, such that eventually all relevant residual transmission must origin from an enhancement via constructive interference through the formation of a supermode. 

\begin{figure}[t]
\includegraphics[width=8cm]{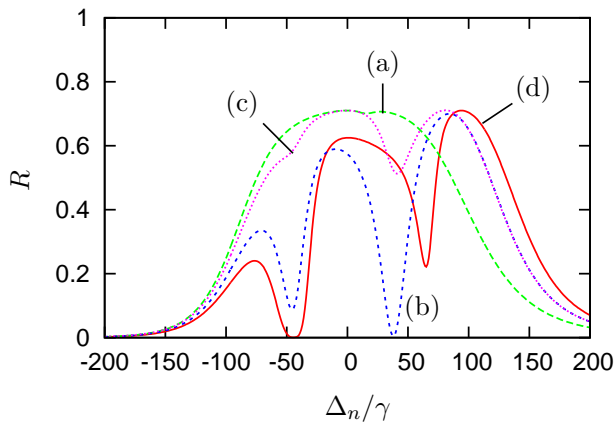}
\caption{\label{asym2}(Color online) Reflection for asymmetric atom-cavity coupling constellations with $N=2$. In (a) no atom couples to the cavities, in (b) one atom couples to the first cavity, in (c) to the second, and in (d) both cavities have nearby atoms. Coupling atoms have $g_{\mc B_i}=70\gamma$, and $L_1=100.3\lambda$. }
\end{figure}

Next, we discuss the reflection properties. Figure~\ref{asym2} shows results for  $R$ for $N=2$ and $L_1=100.3\lambda$. But unlike in Figs.~\ref{Th5} and~\ref{Nxsuper}, we now additionally consider the cases in which only the first cavity couples to an atom ($g_{\mc B_2}=0$), only the second cavity couples to an atom ($g_{\mc B_1}=0$), or both cavities without atom. We observe that the reflection crucially depends on the presence and position of the coupling atoms. For example, if $\Delta_n=0$ and only one atom is located close to cavity 2, no light is transmitted at the first cavity, since $|t(\Delta_n=0)|=0$. Thus no light enters cavity 2 and therefore the dynamics of the coupled chain is totally governed by subsystem 1. However, if the atom is located close to cavity 2, light can be transmitted for $\Delta_n=0$ at cavity 1,  but not at cavity 2. In this case again all light is reflected, but both subsystems influence the reflection. Thus the reflection depends on the coupling position of a single atom. From the example in Fig.~\ref{asym2} we find that if the atom couples to cavity 1, zero reflection can be observed at $\Delta_n\approx 37\gamma$. However, if the atom couples to cavity 2, there is only a slight local minimum at this detuning. Furthermore, the cases with no nearby atom or an atom at both the cavities can be well distinguished  by measuring the reflection intensity, or by comparing the reflection at two frequencies. In an experiment, this behavior could be exploited to determine how many atoms couple to which resonators, which is particularly useful for experiments in which atom-cavity coupling is based on a falling cloud of cold atoms~\cite{turnstile, strong, alton2010}, or if the atoms cannot be trapped reliably over a long time compared to a measurement.

The results for $\Delta T$ and $R$ can be explained in terms of interference between different  pathways the light can take through the system. The contribution to $a_{out,N}$ for the simplest possible pathway $PW1$ in which light is transmitted by all $N$ cavities reads $a_{N,PW1}=a_{in,1}\, \prod_{n=1}^N \,t_n\, \exp[i\phi_n]$.
However, the fully coupled system also allows for extended pathways, e.g., through cavities $1\dots N$, back to $N-1$, and then to the output port via $N$. This pathway contributes $a_{N,PW2}=a_{N,PW1}\,r_N\,r_{N-1}\, \exp(2i\phi_{N-1})$. In case of high $\Delta T$ as for the supermodes presented in Fig.~\ref{Th5}, contributions arising from different pathways interfere constructively. For the pathways PW1  and PW2  the condition for constructive interference is $\phi_{r_N}+\phi_{r_{N-1}}+2\phi_{N-1}=m\cdot 2\pi$ with $m\in\mathbb{Z}$. 

We emphasize that this analysis implies that the order of the subsystems in the chain has no influence on $T$ and thus also $\Delta T$ as long as the input flux $b_{in,N}=0$. In contrast, for the reflection $R$, changing the order of subsystems within the chain in general leads to completely different results, even though the sensitivity depends on the intercavity lengths $L_n$. The reason is that transmitted light necessarily must pass all cavities, whereas reflection can already occur at the first cavity such that the light does not reach the other cavities. Therefore, the determination of the positions at which atoms couple to the chain as in Fig.~\ref{asym2} is only possible via the reflection. In this sense, the reflection provides information about the system properties complementary to those obtained from the transmission.

\begin{figure}[t]
\includegraphics[width=8cm]{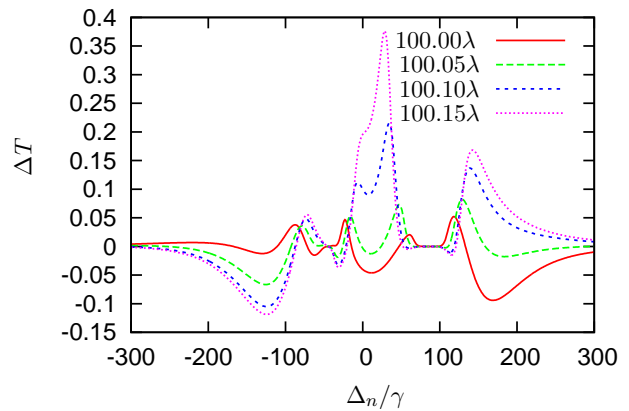}
\caption{\label{super}(Color online) Formation of supermodes as in Fig.~\ref{Th5}, but for $N=3$ subsystems. Parameters are $h_n=50\gamma$, $L_{tot}=200.3\lambda$ and $L_1=100\lambda$, $100.05\lambda$, $100.1\lambda$, and $100.15\lambda$.}
\end{figure}

We conclude with an analysis of the $N=3$ case, which is the simplest realization with two potentially different inter-cavity distances. If at least one cavity in the chain has no nearby atom, the transmission of the coupled system is suppressed in a large range of probe field detunings around $\Delta_n=0$ by the cavity without atom. Then also $\Delta T$ has low values. Since we are interested in supermodes, we focus on the case with atoms. We keep the distance between subsystem 1 and 3 fixed as $L_{tot}=L_1+L_2=200.3\lambda$ and move the central cavity 2, i.e., the distances $L_1$ and $L_2=L_{tot}-L_1$ are varied. In Fig.~\ref{super} we show results for the ``superness'' $\Delta T$ for the parameters $h_n=50\gamma$ and $g_{\mc B_n}=70\gamma$ for different positions of cavity 2. For some detunings only by moving the central cavity-atom subsystem, the transmission changes qualitatively, from a strong supermode character to properties governed by the individual cavities only.  For example, the peak around $\Delta=25\gamma$ has $\Delta T\approx 0.05$ for $L_1=100.0\lambda$, but values up to 0.4 for $L_1=100.15\lambda$. Interestingly, there are also structures which have small supermode character only slightly influenced by $L_1$, e.g., around $\Delta=-50\gamma$.  These results again can be understood from a pathway analysis as before.

In summary, we have analyzed the formation of supermodes in chains of atom-microcavity systems, focusing on the effect of the relative positioning and the chain length on the ``superness'' $\Delta T$. Furthermore, we have shown that in contrast to the transmission, the reflection enables the identification of cavities coupling to an atom, which is crucial for an experimental realization of longer atom-cavity chains.

%


\end{document}